\documentstyle[12pt]{article}
\begin{document}

\hfill UB-ECM-PF 92/11
\mbox{}

\vspace*{1cm}

\begin{center}

{\LARGE \bf
Effective potential and stability of the rigid membrane}

\vspace{8mm}

{\sc E. Elizalde and S.D. Odintsov}\footnote{On leave from
Department of
Mathematics and Physics,
 Pedagogical Institute, 634041 Tomsk, Russia.}
\medskip

Department E.C.M., Faculty of Physics, \\
University of Barcelona, \\
Diagonal 647, 08028 Barcelona, Spain \\
{\it e-mail: eli @ ebubecm1.bitnet}
\vspace{1cm}

{\sl May 1992}

\vspace{1cm}

{\bf Abstract}

\end{center}

The calculation of the effective potential for fixed-end and
toroidal
rigid $p$-branes is performed in the one-loop as well as in the
$1/d$ approximations. The analysis of the involved zeta-functions
(of
inhomogeneous Epstein type) which appear
in the process of regularization is done in full detail.
Assymptotic
formulas (allowing only for exponentially decreasing errors of
order
$\leq 10^{-3}$) are found
which carry all the dependences on the basic parameters of the
theory
explicitly. The behaviour
of the effective potential (specified to the membrane case $p=2$)
is
investigated, and the extrema of this effective potential are
obtained.

\vspace{1cm}

\noindent PACS: \begin{quote} 03.70 Theory of quantized fields, \
11.17
Theories of strings and other extended objects, \  11.10 Field
theory.
\end{quote}

\newpage

\section{Introduction}

The theory of the rigid string [1,2] is interesting because of
the number of its applications to quantum chromodynamics (see
[1-3]
and
references therein) and to statistical physics. Using the same
idea, it is not difficult to construct the action for the
rigid membrane
(or $p$-brane). During the last few years, there has been
some activity in the study of quantum extended objects (like
membranes, see ref. [4] for a review). However, already the
semiclassical quantization of a so highly nonlinear system as a
membrane is a very difficult task [5-7]. Nevertheless, some
interesting issues, like the study of the Casimir energy, the
large-$d$ approximation and the tachyon problem can be addressed
already at the semiclassical level [8,9].

In the present paper, we study  the Casimir energy and the
static potential for the rigid $p$-brane  (at the
classical level, this theory has been considered in ref. [10]),
specifying afterwards our results to the membrane case
($p=2$).
We shall start from the following action, which is
multiplicatively renormalized only in the string case ($p=1$),
\begin{equation}
S= \int d^{p+1} \xi \, \sqrt{g} \left( k + \frac{1}{2\rho^2}
\left[ \Delta (g) X^i \right]^2 \right),
\end{equation}
where $g_{\alpha \beta} = \partial_{\alpha} X^i  \partial_{\beta}
X^i$, $ \alpha=0,1, \ldots, p$, $i=1,2, \ldots, d$, $\Delta (g)
=g^{-1/2}  \partial_{\alpha} g^{1/2} g^{\alpha \beta}
\partial_{\beta} $,  the constant $k$ is the analog of the usual
string tension, and $1/\rho^2$ is the coupling constant
corresponding to the rigid term.

Let us note that the $p$-brane is an interacting system, without
a free part in the action. Hence, one must start from some
classical solution for the ground state, and then study the
quantum fluctuations on such background. In this framework we can
understand how the tachyon appears (if that is the case), if the
background is stable, and also address  some other issues. Owing
to the fact that string theory can be obtained as some
compactification of the membrane [11], we can also expect to find
in this way some new features of string physics.

The paper is organized as follows. In section 2 we calculate the
potential corresponding to two cases: fixed-end and periodic
boundary conditions. In section 3 we obtain the static potential,
that is, the effective potential in the limit of large spacetime
dimensionality. Owing to the difficulty of the exact expressions,
a saddle point analysis is carried out in section 4. In section 5
we provide a short summary of very useful mathematical results on
the inhomogeneous Epstein-type zeta functions and apply them to
the expressions that appear in the process of regularization.
Finally, in section 6 we study the general case and provide some
discussion of the results obtained.

\section{Calculation of the potential}

We consider as the background the classical solutions of the
field equations [8,9] (which are the same for the rigid as for
the usual $p$-brane)
\begin{equation}
X^0_{cl} = \xi_0, \ \ X^{\bot}_{cl} = 0, \ \ X^{d-1}_{cl} =
\xi_1, \ldots,  X^{d-p}_{cl} = \xi_p,
\end{equation}
with  $X^{\bot}_{cl} = (X^1, \ldots , X^{d-p-1})$ and $(\xi_1,
\ldots, \xi_p ) \in {\cal R} \equiv [0,a_1] \times \cdots \times
[0,a_p]$.
We also use the axial gauge of ref. [8], i.e.,
\begin{equation}
X^0 =X^0_{cl}, \ \  X^{d-1} =X^{d-1}_{cl}, \ldots,  X^{d-p}=
X^{d-p}_{cl},
\end{equation}
where the Faddeev-Popov ghosts are absent.

We shall consider the toroidal rigid $p$-brane which has the
boundary
conditions
\begin{equation}
X^{\bot} (0, \xi_1, \ldots, \xi_p) = X^{\bot} (T, \xi_1,
\ldots, \xi_p) =0
\end{equation}
and
\begin{eqnarray}
X^{\bot} (\xi_0,0, \xi_2, \ldots, \xi_p) &=& X^{\bot} (\xi_0,
a_1, \xi_2, \ldots, \xi_p), \nonumber  \\
& \vdots & \\
X^{\bot} (\xi_0, \xi_1, \ldots, \xi_{p-1},0) &=& X^{\bot}
(\xi_0,\xi_1, \ldots, \xi_{p-1}, a_p). \nonumber
\end{eqnarray}
For the fixed-end boundary conditions, equation (4) is exactly
the same, while eqs. (5) are replaced by the following (of
Dirichlet type)
\begin{eqnarray}
X^{\bot} (\xi_0,0, \xi_2, \ldots, \xi_p) &=& \cdots = X^{\bot}
(\xi_0, \xi_1, \ldots, \xi_{p-1},0), \nonumber  \\
X^{\bot} (\xi_0,a_1, \xi_2, \ldots, \xi_p) &=& \cdots =
X^{\bot} (\xi_0, \xi_1, \ldots, \xi_{p-1},a_p).
\end{eqnarray}
The effective potential is given by
\begin{equation}
V= -\lim_{T\rightarrow \infty} \frac{1}{T} \, \ln \int {\cal D}
X^{\bot} \exp (-S).
\end{equation}
Restricting ourselves to the one-loop approximation, we need only
consider the terms which are quadratic in the quantum fields
(this applies to the usual membrane and $p$-brane, see [8,9]).

Integrating out $X^{\bot}$ and using boundary conditions to read
off the resulting $\mbox{Tr} \,  \ln \Delta$ (see [8,9,12]), we get
\begin{eqnarray}
V_{\mbox{fixed end}}& =& k \prod_{i=1}^p a_i + \frac{d-p-1}{2}
\left[ \sum_{n_1, \ldots, n_p=1}^{\infty} \left(
\frac{\pi^2n_1^2}{a_1^2} + \cdots +\frac{\pi^2 n_p^2}{a_p^2}
\right)^{1/2} \right. \nonumber  \\
&+& \left.   \sum_{n_1, \ldots, n_p=1}^{\infty} \left(
\frac{\pi^2n_1^2}{a_1^2} + \cdots +\frac{\pi^2 n_p^2}{a_p^2} + k
\rho^2 \right)^{1/2} \right],
\end{eqnarray}
and
\begin{eqnarray}
V_{\mbox{toroidal}}& =& k \prod_{i=1}^p a_i + \frac{d-p-1}{2}
\left[ \sum_{n_1, \ldots, n_p=-\infty}^{\infty} \left(
\frac{4\pi^2n_1^2}{a_1^2} + \cdots +\frac{4\pi^2 n_p^2}{a_p^2}
\right)^{1/2} \right. \nonumber  \\
&+& \left.   \sum_{n_1, \ldots, n_p=-\infty}^{\infty} \left(
\frac{4\pi^2n_1^2}{a_1^2} + \cdots +\frac{4\pi^2 n_p^2}{a_p^2} +
k \rho^2 \right)^{1/2} \right].
\end{eqnarray}
Observe that the contribution from the higher-derivative mode
appears in (8) and (9) with a positive sign, as it follows from
the path integral. In the second ref. [3] this sign has been taken to be
negative `by hand'. Actually, some arguments based on the
linearized approximations were given there in favor of the
stability of such quantum field theory. Unfortunately, the
expression for the Casimir energy obtained in [3] (second ref.)
 disagrees with the large-$d$ approximation [2].
\section{The limit of large spacetime dimensionality}

We calculate first the static potential ---that is, the effective
potential in the limit of large spacetime
dimensionality.  Such calculation for the usual
Nambu-Goto or Eguchi string  [13] has been carried out in ref.
[14] and for the rigid string in ref. [2].
Let us introduce the composite fields $\sigma_{\alpha \beta}$ for
$\partial_{\alpha} X^{\bot} \cdot \partial_{\beta} X^{\bot} $,
and constrain $\sigma_{\alpha \beta}= \partial_{\alpha} X^{\bot}
\cdot \partial_{\beta} X^{\bot} $ by introducing Lagrange
multipliers $\lambda^{\alpha \beta}$:
\begin{eqnarray}
Z&=& \int {\cal D} X^{\bot} {\cal D} \sigma {\cal D} \lambda \,
\exp
\left\{  -k \int d^{p+1} \xi \, \left[\det (\delta_{\alpha
\beta}
+ \sigma_{\alpha \beta} ) \right]^{1/2} \right. \nonumber  \\
&+& \frac{1}{2\rho^2} \left. \int d^{p+1} \xi
\Delta_0 X^{\bot} \cdot \Delta_0 X^{\bot}
- \frac{k}{2} \int d^{p+1} \xi \lambda^{\alpha \beta}
\left( \partial_{\alpha} X^{\bot} \cdot \partial_{\beta} X^{\bot}
-
\sigma_{\alpha \beta} \right) \right\},
\end{eqnarray}
where $\Delta_0 =\eta^{\alpha \beta} \partial_{\alpha}
\partial_{\beta}$. Integrating over $X^{\bot}$, we get
\begin{equation}
Z= \int  {\cal D} \sigma {\cal D} \lambda \, \exp (-S_{eff}),
\end{equation}
with
\begin{eqnarray}
S_{eff} &=& \frac{1}{2} (d-p-1) \mbox{Tr} \,  \ln \left( \frac{1}{\rho^2}
\Delta_0^2 + k \partial_{\alpha} \lambda^{\alpha \beta}
\partial_{\beta}
\right) \nonumber  \\
&+& kT R^p \left[ (1+\sigma_0)^{1/2} (1+\sigma_1)^{p/2}
-\frac{1}{2}
(\sigma_0 \lambda_0+ p \sigma_1 \lambda_1) \right],
\end{eqnarray}
where we have choosen $a_1= \cdots =a_p=R$, and $\sigma_{\alpha
\beta} =
\mbox{diag} (\sigma_0,\sigma_1, \ldots, \sigma_1)$,
$\lambda_{\alpha
\beta} =
\mbox{diag} (\lambda_0,\lambda_1, \ldots, \lambda_1)$ (compare
with
[8,9], where the case $1/\rho^2=0$ was considered).

In this case we obtain
\begin{eqnarray}
& & \mbox{Tr} \,  \ln  \left[ \Delta_0^2 +k\rho^2 (\lambda_0 \partial_0^2+
\lambda_1 \vec{\partial}_x^2 ) \right] \nonumber  \\
&=& \mbox{Tr} \,  \ln \left[ (\partial_0^2 + \vec{\partial}_x^2)^2 +k\rho^2
(\lambda_0
\partial_0^2+ \lambda_1 \vec{\partial}_x^2 ) \right] \nonumber
\\
&=& \mbox{Tr} \,  \ln \left\{ \left[ \partial_0^2 + \left(
\vec{\partial}_x^2
+\frac{k\rho^2 \lambda_0}{2} \right)-\sqrt{ k\rho^2 (\lambda_0
-\lambda_1) \vec{\partial}_x^2+ \frac{k^2 \rho^4\lambda_0^2}{4}}
\,
\right]
\right. \\
&\times & \left. \left[ \partial_0^2 + \left(
\vec{\partial}_x^2
+\frac{k\rho^2 \lambda_0}{2} \right)+\sqrt{ k\rho^2 (\lambda_0
-\lambda_1) \vec{\partial}_x^2+ \frac{k^2 \rho^4\lambda_0^2}{4}}
\,
\right]
\right\}. \nonumber
\end{eqnarray}
The spectrum for the boundary conditions (2)-(6) is known. Using
this
spectrum and evaluating the Trln terms by means of analytic
regularization for large $T$, we obtain (see [8,9] for details of
this
method)
\begin{eqnarray}
S_{eff} &=&
 kT R^p \left\{ (1+\sigma_0)^{1/2} (1+\sigma_1)^{p/2}
-\frac{1}{2}
\right. (\sigma_0 \lambda_0+ p \sigma_1 \lambda_1)  \nonumber  \\
&+&  \frac{d-p-1}{2kR^{p+1}} \left[ \sum_{\vec{n}} \left(
\pi^2 \vec{n}^2
+ \frac{k\rho^2 \lambda_0 R^2}{2} - \sqrt{ k\rho^2 (\lambda_0
-\lambda_1) \vec{n}^2 R^2 + \frac{k^2\rho^4 \lambda_0^2 R^4}{4}}
\right)^{1/2} \right. \label{sef1}  \\
&+&  \left. \left.  \sum_{\vec{n}} \left( \pi^2 \vec{n}^2
+ \frac{k\rho^2 \lambda_0 R^2}{2} + \sqrt{ k\rho^2 (\lambda_0
-\lambda_1) \vec{n}^2 R^2 + \frac{k^2\rho^4 \lambda_0^2 R^4}{4}}
\right)^{1/2} \right] \right\}, \nonumber
 \end{eqnarray}
where for the fixed-end $p$-brane
 $\vec{n}^2 =n_1^2+ \cdots + n_p^2$ and
$\sum_{\vec{n}}$ means $\sum_{n_1, \ldots, n_p=1}^{\infty}$ as in
(8),
while for the toroidal $p$-brane
 $\vec{n}^2 =4(n_1^2+ \cdots + n_p^2)$ and
$\sum_{\vec{n}}$ means $\sum_{n_1, \ldots, n_p=-\infty}^{\infty}$
as in
(9).

\section{A saddle point analysis}

The functions that appear on the right hand side of (\ref{sef1})
are
rather complicated to analyze. To our knowledge, they have never
been
considered
 in the literature and will be the object of a separate
investigation. (Note that in the case of
the usual Dirac $p$-brane [8,9] these functions are simply
constants).
So we shall have here little to say about the corresponding
effective
potential, only, for example, that as $R\rightarrow
\infty$, $V\sim V_{cl} =kR^p$.
Rewriting the expression for the static potential identically as
\begin{eqnarray}
V &=&
 k R^p \left[ (1+\sigma_0)^{1/2} (1+\sigma_1)^{p/2} -\frac{1}{2}
\right. (\sigma_0 \lambda_0+ p \sigma_1 \lambda_1)  \nonumber  \\
&+&  \left. \frac{d-p-1}{2kR^{p+1}}
 K(k\rho^2 R^2, \lambda_0, \lambda_1)
\right],
\end{eqnarray}
we are led to the four saddle-point equations:
\begin{eqnarray}
\lambda_0 &=& (1+\sigma_1)^{p/2} (1+\sigma_0)^{-1/2}, \nonumber
\\
\lambda_1 &=& (1+\sigma_1)^{p/2-1} (1+\sigma_0)^{1/2}, \nonumber
\\
\sigma_0 &=& \frac{d-p-1}{kR^{p+1}} \, \frac{\partial
 K(k\rho^2 R^2, \lambda_0, \lambda_1)}{\partial \lambda_0}, \\
\sigma_1 &=& \frac{d-p-1}{kpR^{p+1}} \, \frac{\partial
 K(k\rho^2 R^2, \lambda_0, \lambda_1)}{\partial \lambda_1},
\nonumber
\end{eqnarray}
By eliminating from these equations $\sigma_0$ and $\sigma_1$, we
get
\begin{eqnarray}
(\lambda_1\lambda_0)^{p/(p-1)} \lambda_0^{-2}-1  &=&
\frac{d-p-1}{kR^{p+1}} \, \frac{\partial
 K(k\rho^2 R^2, \lambda_0, \lambda_1)}{\partial \lambda_0},
\nonumber  \\
(\lambda_1\lambda_0)^{1/(p-1)} -1  &=&
 \frac{d-p-1}{kpR^{p+1}} \, \frac{\partial
 K(k\rho^2 R^2, \lambda_0, \lambda_1)}{\partial \lambda_1}.
\end{eqnarray}
In principle, if not analytically, it is of course possible to
eliminate (let us say) $\lambda_0$ from these two equations by
means of
a numerical calculation, and to rewrite $V$ in terms of
$\lambda_1$
only. After doing this, one can study $V$ as a function of $R$
and
in
terms of this parameter $\lambda_1$, in order to see if the
tachyon
is
in the spectrum. We do not do this work here, since our purpose
is
just to show the possibility, in principle, of calculating the
static
potential for a rigid string. However, we are ready to look to
some
interesting limiting
case
of these equations. Let us assume that $\lambda_0=\lambda_1
\equiv
\lambda$ (such a choice has been taken for the rigid string in
[2]).
Then
\begin{eqnarray}
V &=&
 k R^p \left\{ (1+\sigma_0)^{1/2} (1+\sigma_1)^{p/2} -\frac{1}{2}
\right. \lambda (\sigma_0 + p \sigma_1)  \nonumber  \\
&+&  \left. \frac{(d-p-1)\pi}{2kR^{p+1}} \left[  \sum_{\vec{n}}
\sqrt{ \vec{n}^2} +  \sum_{\vec{n}} (\vec{n}^2 + k(\rho/\pi)^2
\lambda
R^2)^{1/2} \right] \right\}.
\end{eqnarray}
It follows from the saddle point equations that $\sigma_0
=\sigma_1$,
$\lambda =(1+\sigma)^{(p-1)/2}$, and
\begin{equation}
V=kR^p \left[ \lambda^{(p+1)/(p-1)} - \frac{p+1}{2} \lambda
(\lambda^{2/(p-1)} -1) +
 \frac{(d-p-1)\pi}{2kR^{p+1}}  K(k\rho^2 \lambda R^2) \right].
\end{equation}
Here we have defined
\begin{equation}
K(k\rho^2 \lambda R^2) \equiv   \sum_{\vec{n}} \sqrt{
\vec{n}^2} + \sum_{\vec{n}} (\vec{n}^2 + k (\rho/\pi)^2 \lambda
R^2)^{1/2},
\end{equation}
and the last saddle point equation gives
\begin{equation}
-\frac{p+1}{2} \lambda^{2/(p-1)}
+ \frac{(d-p-1)\pi}{2kR^{p+1}} \, \frac{\partial
 K(k\rho^2 \lambda R^2)}{\partial \lambda} =0.
\end{equation}

\section{Explicit expressions for the inhomogeneous Epstein
functions}

The expressions to be regularized are (in general) of the
inhomogeneous Epstein form \cite{e1}
\begin{equation}
E_p^c(s)\equiv \sum_{n_1, \ldots, n_p=1}^{\infty} (n_1^2+ \cdots
+
n_p^2+c^2)^{-s}, \label{ehp}
\end{equation}
allowing for $c= 0$. These functions are not easy to deal with
(for $p>1$, the case $p=1$ is the only one that has been
investigated in
the
literature) and refs. \cite{e1} can be considered as pioneering
in this respect. There, very general formulas have been derived
for
the functions
\begin{equation}
M_N^c(s;a_1, \ldots, a_N; \alpha_1, \ldots, \alpha_N) \equiv
\sum_{n_1, \ldots, n_N=1}^{\infty} (a_1n_1^{\alpha_1}+ \cdots +
a_Nn_N^{\alpha_N} +c^2)^{-s},
\end{equation}
for any value of $c$, such as
\begin{eqnarray}
&& M_N^c  (s;a_1, \ldots, a_N; \alpha_1, \ldots, \alpha_N)  =
\frac{1}{a_N^s \Gamma (s)} \sum_{p=0}^{N-1} \sum_{C_{N-1,p}}
\prod_{r=1}^p \frac{b_{i_r}^{-1/\alpha_{i_r}}}{\alpha_{i_r}}
\Gamma
\left( \frac{1}{\alpha_{i_r}} \right) \nonumber  \\ &\times&
\sum_{k_{j_1},
\ldots, k_{j_{N-p-1}}=0}^{\infty} \Gamma \left( s+
\sum_{l=1}^{N-p-1}
k_{j_l}- \sum_{r=1}^p \frac{1}{\alpha_{i_r}} \right)  \label{egf}
\\
& \times & \prod_{l=1}^{N-p-1} \left[ \frac{(-
b_{j_l})^{k_{j_l}}}{k_{j_l}!} \zeta (-\alpha_{j_l} k_{j_l} )
\right] \zeta \left( \alpha_n \left( s+  \sum_{l=1}^{N-p-1}
k_{j_l}- \sum_{r=1}^p \frac{1}{\alpha_{i_r}} \right) \right) +
\Delta_{ER}, \nonumber
\end{eqnarray}
with $b_{i_r} \equiv a_{i_r}/a_N$ (notice the slight erratum in
eq.
(3.22) of the second ref. \cite{e1}), and $1\leq i_1 < \cdots <
i_p \leq
N-1$, $1\leq j_1 < \cdots < j_{N-p-1} \leq N-1$, being $i_1,
\ldots,
i_p,j_1, \ldots, j_{N-p-1}$ a permutation   of $1,2,
\ldots, N-1$. The sum on $C_{N-1,p}$ means sum over the
$\left(_{\
\
p}^{N-1} \right)$ choices of the indices $i_1, \ldots, i_p$ among
the $1,2, \ldots, N-1$.

 In our case, $a_1= \cdots =a_p=1$ and
$\alpha_1 = \cdots =\alpha_p=2$, and the involved general formula
above,  (\ref{egf}), simplifies considerably.  We have, for
$c=0$,
\begin{equation}
E_p(s) = \frac{(-1)^{p-1}}{2^{p-1}} \frac{1}{\Gamma (s)}
\sum_{j=0}^{p-1} (-1)^j \left(_{ \ \, j}^{p-1} \right) \Gamma
\left(
2s- j \right) \zeta \left(s- \frac{j}{2} \right)+ \Delta_{ER},
\label{cic}
\end{equation}
and for $c\neq 0$,
\begin{equation}
E_p^c(s) = \frac{(-1)^{p-1}}{2^{p-1}} \frac{1}{\Gamma (s)}
\sum_{j=0}^{p-1} (-1)^j \left(_{ \ \, j}^{p-1} \right) \Gamma
\left(
s- \frac{j}{2} \right) E_1^c \left(s- \frac{j}{2} \right) +
\Delta_{ER}.
\label{cmc}
\end{equation}
Notice that the poles of this function arise from those of
$E_1^c(s-j/2)$, which are obtained for the values of $s$ such
that
$s-j/2=1/2,-1/2,-3/2, \ldots$. They  are poles of order one at
$s=p/2,(p-1)/2, p/2-1, \ldots$, except for $s=0,-1,-2, \ldots$,
then the function is finite (owing to the $\Gamma (s)$ in the
denominator). These poles are removed by zeta-function
regularization \cite{er2}.

Alternatively, a very useful and exact recurrent formula is the
following \cite{e1}
\begin{eqnarray}
&& E_N^c(s;a_1, \ldots, a_N) \equiv
\sum_{n_1, \ldots, n_N=1}^{\infty} (a_1n_1^2+ \cdots +
a_Nn_N^2 +c^2)^{-s} \nonumber  \\ &=& - \frac{1}{2}
E_{N-1}^c(s;a_2, \ldots,
a_N)
+ \frac{1}{2} \sqrt{\frac{\pi}{a_1}} \frac{\Gamma (s-1/2)}{\Gamma
(s)}  E_{N-1}^c(s-1/2;a_2, \ldots, a_N) \label{enc1} \\ &+&
\frac{\pi^s}{\Gamma
(s)} a_1^{-s/2} \sum_{k=0}^{\infty} \frac{a_1^{k/2}}{k!(16\pi)^k}
\prod_{j=1}^k [(2s-1)^2-(2j-1)^2]
 \sum_{n_1, s, n_N=1}^{\infty} n_1^{s-k-1} \nonumber  \\ & \times
& (a_2n_2^2 +
\cdots + a_Nn_N^2 +c^2)^{-(s+k)/2} \exp \left[ - \frac{2
\pi}{\sqrt{a_1}} n_1 (a_2n_2^2+ \cdots + a_Nn_N^2 +c^2)^{1/2}
\right]. \nonumber
\end{eqnarray}
The recurrence starts from expression
\begin{eqnarray}
E_1^c(s;1) &=& -\frac{c^{-2s}}{2} + \frac{\sqrt{\pi}}{2}
\frac{\Gamma (s-1/2)}{\Gamma (s)} c^{-2s+1} \nonumber  \\
&+& \frac{2\pi^s c^{-s+1/2}}{\Gamma (s)} \sum_{n=1}^{\infty}
n^{s-
1/2} K_{s-1/2} (2\pi n c). \label{e1c1}
\end{eqnarray}

 In order to deal with the derivative of the
function $K$ above, one can follow two equivalent procedures:
either do first the usual analytic continuation, and then take
$s=-
1/2$ and the derivative afterwards, or else take first the
derivative of  (\ref{ehp}), perform the analytic continuation and
put $s=+1/2$ at the end. The result is exactly the same. In
either
way other non-trivial series commutations have to be performed
(see \cite{er3} and references therein). We get, in particular,
for $c\neq 0$
\begin{eqnarray}
E_2^c(s)&=&-\frac{1}{2} E_1^c(s) + \frac{\sqrt{\pi}}{2} \,
\frac{\Gamma \left(s-\frac{1}{2} \right)}{\Gamma (s)} E_1^c
\left(
s-\frac{1}{2} \right) + \Delta_{ER}, \\
E_3^c(s)&=&\frac{1}{4} E_1^c(s) - \frac{\sqrt{\pi}}{2} \,
\frac{\Gamma \left(s-\frac{1}{2} \right)}{\Gamma (s)} E_1^c
\left(
s-\frac{1}{2} \right) + \frac{\pi}{4(s-1)} E_1^c \left(
s-1\right)
+ \Delta_{ER}, \nonumber
\end{eqnarray}
and similar expressions for $c=0$, as it is clear from
(\ref{cic})
and (\ref{cmc}) (it has been proven in \cite{er2} that this case
can be obtained from the former by analytically continuing in the
parameter $c$). Remember that $\Delta_{ER}$ is the well-known
term
(here exponentially
small), which was found in \cite{er3}.
As our final interest is numerical approximation (see above), we
will not take into account exponentially small terms (let us
point
out
that these expansions are asymptotic and very quickly convergent)
\cite{e1}.
Notice, moreover, that for $c=0$ there is no dependence on
$\lambda$, so that the corresponding term does not contribute to
eq. (22).

{}From the above expressions, for any value of $p$ there is no
difficulty in obtaining the value of $\lambda$ which solves (22).
In particular, for $p=2$ we have:
\begin{equation}
\lambda \simeq \left(\frac{3-d}{12\pi}\right)^{2/3} k^{1/3}
\rho^2
\end{equation}
this is a sensible root for $k\cdot \rho$ big (specifically, for
$1/R << k^{2/3} \rho^2$) and $d\neq 3$. For
$p=3$ the result is:
\begin{equation}
\lambda \simeq \frac{1}{2^{10}} \left[ \frac{3(4-d)}{ \pi}
\right]^2 \frac{k \rho^6}{R^2},
\end{equation}
which is a sensible root for $k\cdot \rho$ big (specifically, for
$ k \rho^4 >>1$) and $d\neq 4$.
Let us now substitute these values into the expression of $V$,
and
look for the derivative $\partial V / \partial R$. We get, for
$p=2$, an
expression of the form
\begin{equation}
\frac{\partial V_2}{\partial R} = c_1 R+ \frac{c_{-1}}{R} +
\frac{c_{-2}}{R^2},
\end{equation}
which has always one real root (at least), $R_2$. It
corresponds
to a minimum of $V$ for $d>3$ and reasonable values of the
constants
involved. Substituting back into $V(R)$ we see that the minimum
is
attained at
\begin{equation}
V_2(R_2)= kR_2^2 \left[ 3(1+\sigma_0)^{1/2} (1+\sigma_1)
-\frac{3}{2}
\left( \frac{3-d}{12\pi} \right)^{2/3} (\sigma_0+2\sigma_1)
k^{1/3}
\rho^2+ \frac{(d-3)^2}{48\pi^2} \rho^6 \right],
\end{equation}
i.e., for $k$ big it is obtained at  a negative value of $V$,
while
for $\rho$ big it is reached at a positive value of $V$.

The case $p=3$ is quite different. We get then
\begin{equation}
 V_3 ( R) = \alpha_1 R^3- \alpha_2 R+ \frac{\beta}{R},
\end{equation}
so that its derivative has two real roots
\begin{equation}
R_{\pm}=\pm \left[ \frac{\alpha_2+(\alpha_2^2+12\alpha
\beta)^{1/2}}{6\alpha_1} \right]^{1/2},
\end{equation}
one of which is seen to correspond to a maximum and the other to
a
minimum of $V$. Moreover, two additional, complex roots appear.
The
minimum for $V$ is now attained at a negative value of $V$ when
either
$k$ or $\rho$ are big enough and, conversely, at a positive value
of the
potential for $k$ or $\rho$ small.
Note that in order to find the critical radius at which the potential
becomes complex (so that the static approximation breaks down and
tachyons appear) it is necessary to do the analysis with the general
expression (15) directly. One can conjecture from our results here that
the rigid membrane is  tachyon-free (no critical radius exists),
as it is also the case with rigid strings (see the second ref. [2]
and the first ref. [3]). At least for the limiting situation
discussed above, this is in fact the case.

\section{Discussion of the general case}

Having done the calculation for this special case, corresponding
to
the limit of large
spacetime dimensionality, and armed with the full equation
(\ref{egf}), we can now be more ambitious and consider the
one-loop
effective potential, (8)-(9), without further limit or
approximation. For the  sake of conciseness, we shall restrict
ourselves to  $p=2$ ---but it is obvious that we could consider
as
well any other value of $p$. We rely on equations (\ref{enc1})
and
(\ref{e1c1}), which specialized to $p=2$ yield, after some work,
\begin{eqnarray}
&& \sum_{n_1,n_2=1}^{\infty} \sqrt{
\left(\frac{n_1}{a_1} \right)^2 +
\left(\frac{n_2}{a_2} \right)^2}
= \frac{1}{24} \left( \frac{1}{a_1} + \frac{1}{a_2} \right) -
\frac{\zeta (3)}{8\pi^2}
 \left( \frac{a_1}{a_2^2} + \frac{a_2}{a_1^2} \right) \nonumber
\\
&-& \frac{\pi^{3/2}}{2\sqrt{a_1a_2}} \left[ \exp \left( -2\pi
\frac{a_1}{a_2} \right) \left( 1+ {\cal O} (10^{-3}) \right)
\right],
\end{eqnarray}
and (this one after additional regularization, see above)
\begin{eqnarray}
&&\sum_{n_1,n_2=1}^{\infty} \sqrt{
\left(\frac{n_1}{a_1} \right)^2 +
\left(\frac{n_2}{a_2} \right)^2 +c^2}
= \frac{c}{4} - \frac{\pi}{6} a_1a_2c^3 \nonumber  \\
&+& \left( \frac{1}{4\pi} \sqrt{\frac{c}{a_2}} -\frac{c a_1}{4\pi
a_2}
\right) \left[ \exp \left( -2\pi ca_2
 \right) \left( 1+ {\cal O} (10^{-3}) \right) \right].
\end{eqnarray}
In both cases we have assumed (this is, of course, no
restriction) that
$a_2\leq a_1$.

These expressions are really valuable. They are asymptotic, the
last
term (already of exponential kind) being of order $10^{-3}$ with
respect
to the two first ones, and the not explicitly written
contributions
being of order $10^{-6}$. To our knowledge, the second expression
---which can be termed
as of inhomogeneous Epstein type--- has never been discussed in
the
literature.

 For fixed-end boundary conditions
and not taking into account exponentially-small terms, we obtain
\begin{eqnarray}
V_{f.e.} &\simeq & ka_1a_2 + \frac{(d-3)\pi}{24} \left[
\frac{1}{2}
 \left( \frac{1}{a_1} + \frac{1}{a_2} \right) -
\frac{3\zeta (3)}{2\pi^2}
 \left( \frac{a_1}{a_2^2} + \frac{a_2}{a_1^2} \right) \right.
\nonumber  \\
&+& \left. \frac{3}{\pi} \sqrt{k} \rho - \frac{2}{\pi^2} k^{3/2}
\rho^3 a_1a_2 \right]. \label{vfe}
\end{eqnarray}
It is now straightforward to perform the analysis of extrema of
$V$. For
brevity,
we shall only discuss here some particular cases. First, the one
which
is obtained
from the two Lagrange equations for the extrema of $V$ as a
function of
$a_1$ and $a_2$ only, for $a_1=a_2\equiv a$, which is reached for
 \begin{equation}
a= \left( \frac{\frac{3-d}{8} [3 \zeta (3)-\pi^2]}{12\pi k +
(3-d)k^{3/2} \rho^3} \right)^{1/3}.
\end{equation}
It can be seen that for $12\pi k +(3-d) k^{3/2} \rho^3 <0$ this
point is
a minimum. On the contrary, it is a maximum for
 $12\pi k +(3-d) k^{3/2} \rho^3> 0$. Consistency with the range
of
validity of the series expansion above is obtained for
\begin{equation}
2\pi \sqrt{k} \rho > \frac{6}{\rho^2} >>1.
\end{equation}
typical values for which this is valid are: $\rho \simeq 2/3$, $k
\simeq
4$, $2\pi c \simeq 8$.

Keeping now $a_1$ and $a_2$ fixed (but arbitrary), we see that
(for
$d>3$) in terms of $\rho$, $V$ is unbounded from below, being
always
negative for $\rho$ big enough. Considering $V$ as a function
of $k$, the situation is similar. Finally, in a sense the
analysis of
ref. [8] is
still valid here, when we fix the values of $k$, $\rho$ and of
the area
$A=a_1a_2$: the minima of the potential are obtained for
elongated
(stringy) membranes ($a_1 / a_2 $ small). Notice, however, that
even for the particular case considered in [8], our asymptotic
expansion
provides a more universal expression, because it is valid for any
value of $a_2 \leq a_1$ (this is again not restrictive, in the
end).
It also goes without saying that, from our general formula
(\ref{vfe}) for the potential $V= V(a_1,a_2,k,\rho,d)$, one can
perform
a simultaneous analysis on {\it all} the different parameters at
the
same time ---e.g. in order to look for local minima of the
potential
hypersurface--- the explicit dependences on $k$ and $\rho$ being
also
basic contributions of the present work.

In the case of toroidal boundary conditions, again neglecting
exponentially-small contributions, we get (for a very
detailed
discussion
of the relations between the different boundary conditions see
the last
of references \cite{e1})
\begin{equation}
V_{tor} \simeq  ka_1a_2 + \frac{(d-3)\pi}{2} \left[
-\frac{\zeta (3)}{\pi^2}
 \left( \frac{a_1}{a_2^2} + \frac{a_2}{a_1^2} \right)
  \frac{1}{\pi} \sqrt{k} \rho  - \frac{1}{6\pi^2} k^{3/2}
\rho^3  a_1a_2 \right].
\end{equation}
The particular extremum for $a_1=a_2 \equiv a$ is a minimum of
$V$
provided that $\sqrt{k} \rho^3 > 12 \pi$ (it is a maximum for
 $\sqrt{k} \rho^3 < 12 \pi$). Consistency with the series
expansion
implies now
\begin{equation}
 \sqrt{k} \rho > \frac{12\pi}{\rho^2} >>1.
\end{equation}
which can be met
typically for values of $\rho \simeq 3$, $k
\simeq 9$, $2\pi c \simeq 9$ ---but, of course, as in the former
case,
the range of allowed values is much widder.

\vspace{2cm}

\noindent{\large \bf Acknowledgments}

We would like to thank Profs. Dom
nec Espriu, Ulf Lindstrm,
Kelly
Stelle and Robin Tucker for fruitfull discussions on connected
problems.
 S.D.O. thanks the members of the Department E.C.M. of
Barcelona University for the kind hospitality, and E.E. the
Alexander von Humboldt Foundation for continued attentions.
This work has
been supported by Direcci"n General de Investigaci"n
Cient!fica y Tcnica (DGICYT), research projects
 PB90-0022 and SAB92-0072.

\end{document}